\documentstyle[12pt,iopconf,epsfig]{article}
\begin{document}
\title{Big Bang Nucleosynthesis: Current Status}

\author{Gary Steigman\footnote{E-mail : steigman@mps.ohio-state.edu.}}

\affil{Departments of Physics and Astronomy, The Ohio State 
University, 174 West 18th Avenue, Columbus, OH 43210, USA}

\beginabstract
During its hot, dense, early evolution the Universe was a primordial 
nuclear reactor, synthesizing the light nuclides D, $^3$He, $^4$He 
and $^7$Li in the first thousand seconds.  The presently observed 
abundances of these relic nuclides provide a unique window on the 
early Universe.  The implications of current observations for cosmology 
(the universal density of nucleons) and for particle physics (new 
particles beyond the standard model) will be reviewed.  The present 
data appear to be in rough agreement with the predictions of the 
standard, hot, big bang model for three species of light neutrinos,
and a nucleon-to-photon ratio restricted to a narrow range of 
3-4 parts in 10 billion.  On closer inspection, however, a tension 
is revealed between the inferred primordial abundances of deuterium 
and helium-4.  Although observations of deuterium in nearly primordial, 
high-redshift QSO absorbers may help to relieve this tension, current 
data appear to exacerbate the crisis.  Resolution of this conflict may 
lie with the data (statistical uncertainties?), with the analysis of 
the data (systematic uncertainties?), or with the fundamental physics 
(massive, unstable, and/or degenerate neutrinos?).  Independent (non-BBN) 
evidence from cosmological observations of large scale dynamics and 
structure may be useful in deciding among the current options.
\endabstract

\section{Introduction}

During the first thousand seconds in the evolution of the Universe, as 
it expanded and cooled from very high densities and temperatures, nuclear 
reactions transformed neutrons and protons into astrophysically interesting 
abundances of the light nuclides deuterium, helium-3, helium-4 and lithium-7.  
In the context of Standard, Big Bang Nucleosynthesis (SBBN; homogeneous, 
isotropic expansion, three flavors of non-degenerate neutrinos) these 
abundances depend on only one adjustable parameter, the nucleon density.  
Since as the Universe expands all densities decrease, it is useful to 
express the nucleon density in terms of a nearly constant parameter, 
the ratio of nucleons to photons, which has barely changed at all since 
the annihilation of electron-positron pairs in the early Universe.

\begin{equation}
\eta \equiv n_{\rm N}/n_{\gamma} \ \ ; \ \ \eta_{10} \equiv 10^{10}\eta
\end{equation}
The contribution of nucleons (baryons) to the universal mass-density 
may be written as the dimensionless ratio of the baryon density to 
the critical density (which depends on the present value of the Hubble 
parameter: H$_{0} = 100\,h\,$kms$^{-1}$Mpc$^{-1}$; $\Omega_{\rm B} \equiv 
\rho_{\rm B}/\rho_{crit}$).

\begin{equation}
\Omega_{\rm B}\,h^{2} = \eta_{10}/273
\end{equation}
SBBN is an overdetermined theory in that the observable abundances 
of four nuclides are predicted on the basis of one free parameter.  
In Figure 1 the predictions of the primordial abundances are shown 
for a wide range of $\eta$.  SBBN is falsifiable in that it is possible 
that {\bf no} value of $\eta$ will be consistent with the primordial 
abundances inferred from the observational data.  Furthermore, consistency 
requires that {\bf if} an acceptable value of $\eta$ is found, the 
corresponding nucleon density at present, $\Omega_{\rm B}$, is in agreement 
with other astronomical observations.  Indeed, since there must be enough 
baryons to account for the visible matter in the Universe, but not too 
many to violate constraints on the total mass density, the {\it interesting} 
range of $\eta$ in Figure 1 is restricted to $3\times 10^{-11} - 1\times 
10^{-8}$.  Even so, note the enormous range in the predicted abundances 
of deuterium and lithium.  Over this same range in $\eta$ the predicted 
primordial mass fraction of $^4$He, Y$_{\rm P}$, hardly changes at all.  
As we shall soon see, consistency between D and $^4$He provides a key 
test of SBBN.

\begin{figure}
\centerline{\psfig{file=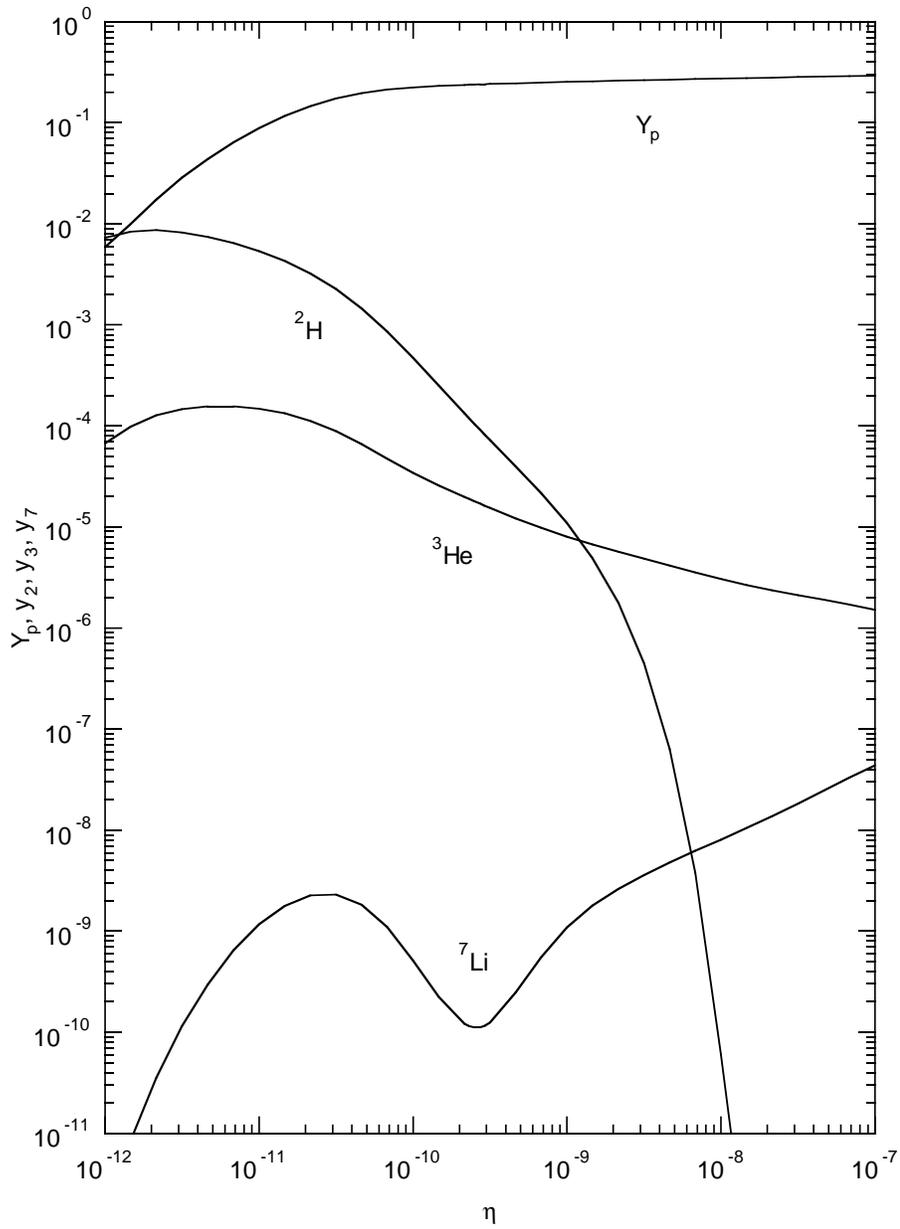,width=0.9\textwidth}}
\vspace{-24pt}
\caption{SBBN-predicted abundances of the light nuclides versus $\eta$.  
The $^4$He mass fraction (Y$_{\rm P}$) is shown along with the ratio by 
number to hydrogen of D ($^2$H, $y_2$), $^3$He ($y_3$), and $^7$Li ($y_7$).
This figure is from D. Thomas.}
\end{figure}

\subsection{Status Quo Ante}

SBBN has provided one of the most spectacular confirmations of the 
standard, hot Big Bang model of cosmology.  Along with the Hubble 
expansion and the cosmic background radiation, SBBN is one of the 
pillars of the standard model.  It is the only one offerring a connection 
between particle physics and cosmology.  For example, Walker \etal 
(1991) reanalyzed the relevant observational data to make a critical 
confrontation between predictions and observations.  Walker \etal (1991) 
concluded that SBBN was consistent with the observational data for 
$\eta_{10} = 3.4\pm0.3$ ($\Omega_{\rm B}h^{2} \approx 0.01$), making 
the nucleon density one of the very best determined of all cosmological 
parameters.  Furthermore, they noted that to preserve this consistency 
required that the total number of ``equivalent", light neutrinos 
(particles which were relativistic at BBN), N$_{\nu}$, should not 
exceed 3.4.  With the three known flavors of neutrinos (provided none 
has a mass comparable to MeV energies), this leaves very little room 
for any new (light) particles ``beyond the standard model".  At this 
point it may have been tempting to declare victory for SBBN and to move 
on to other problems in cosmology.  However, it was still important 
to subject the standard model to ever more precise observational tests 
in order to reaffirm its consistency and to narrow even further the 
bounds on the nucleon density and on particle physics beyond the standard 
model.  To our surprise, my colleagues and I found a dark cloud looming 
on the horizon of the standard model (Hata \etal 1995).

\subsection{A Crisis For SBBN?}
There had, in fact, always been a ``tension" between the predictions of 
SBBN and the inferred primordial abundances of D and $^4$He (Kernan \& 
Krauss 1994, Olive \& Steigman 1995) in the sense that while deuterium 
favored ``high" values of $\eta$ (Steigman \& Tosi 1992, 1995), helium-4 
pointed towards lower values (Olive \& Steigman 1995).  Indeed, in a 
reanalysis focusing on the $^4$He abundance, Olive \& Steigman (1995) 
found for the best estimate of the number of equivalent light neutrinos, 
N$_{\nu} = 2.2$.  Only a generous error estimate permitted consistency 
with SBBN.  It was, therefore, not entirely unexpected when Hata \etal 
(1995) identified a ``crisis" for SBBN in their comparison of the best 
estimates of the primordial abundances derived from the observational 
data with those predicted by SBBN.  The problem is illustrated in Figure 
2 which concentrates on the key nuclides, D, $^4$He and $^7$Li.  While 
the $^4$He abundance is just barely consistent with the low end of the 
$\eta$ range identified by Walker \etal (1991), the deuterium abundance 
is only consistent with the upper end of that range.  Note that due to 
its ``valley" shape and to the relatively larger uncertainties in its 
predicted and inferred abundances, lithium is consistent with either 
deuterium or helium.  Since it thus fails to discriminate between the 
low $\eta$ favored by helium and the higher $\eta$ preferred by deuterium, 
lithium is ignored in the following discussion.
\begin{figure}
\centerline{\psfig{file=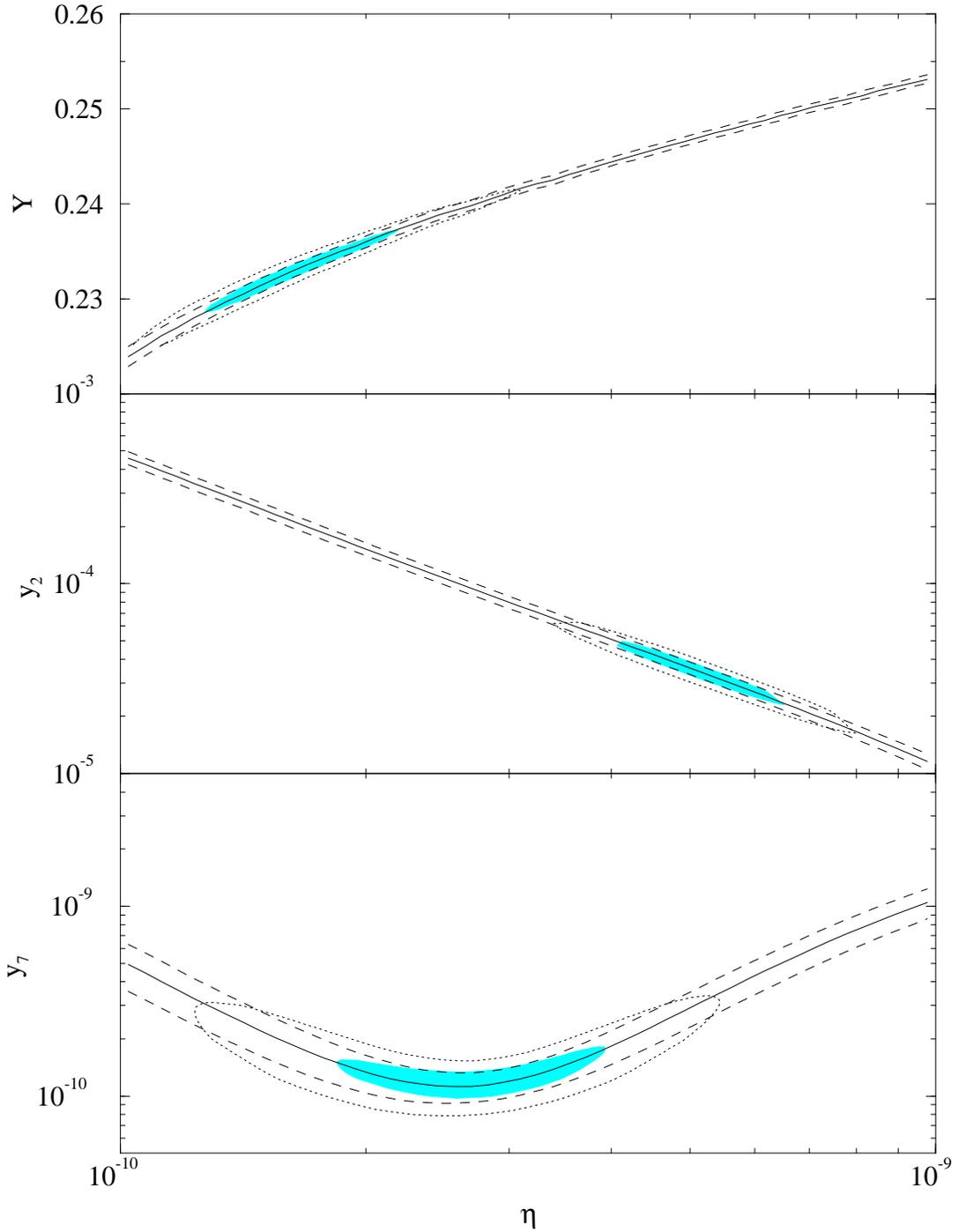,width=0.9\textwidth}}
\caption{SBBN predictions (solid lines) for $^4$He (Y), D ($y_2$), and 
$^7$Li ($y_7$) with the theoretical uncertainties (1$\sigma$) estimated 
by the Monte Carlo method (dashed lines).  Also shown are the regions 
constrained by the observations at 68\% and 95\% C.L. (shaded regions 
and dotted lines, respectively).  This figure is from Hata \etal (1995).}
\end{figure}

Three possible resolutions of the challenge to SBBN posed by the D -- $^4$He 
conflict suggest themselves.  Perhaps the primordial abundance of helium 
inferred from observations of extragalactic \hii regions (see, \eg, Olive \& 
Steigman 1995 and Olive, Skillman, \& Steigman 1997) is too small (see, 
\eg, Izotov, Thuan, \& Lipovetsky 1994 and Izotov \& Thuan 1997).  If 
the primordial helium mass fraction were closer to 0.25 than to 0.23, 
the challenge to SBBN evaporates.  Since several dozen \hii regions are 
observed, the statistical uncertainty in Y is small, typically $\pm 0.003$ 
or smaller (Olive, Skillman, \& Steigman 1997, Izotov \& Thuan 1997).  
But systematic errors, such as those due to uncertainties in the corrections 
for unseen neutral helium, for collisional ionization, for temperature 
fluctuations and, especially, for underlying stellar absorption, may well 
be much larger.  Alternatively, it could be that our adopted primordial 
deuterium abundance is too small.  If the true primordial ratio (by number) 
of deuterium to hydrogen were a few parts in $10^4$ rather than the few 
parts in $10^5$ inferred from observations in the solar system and the 
local interstellar medium (ISM), lower $\eta$, consistent with Y$_{\rm P}$, 
is allowed (see Fig. 2).  This local estimate of the deuterium abundance 
requires an extrapolation from ``here and now" (solar system, ISM) to 
``there and then" (primordial).  Any errors in this extrapolation open 
the door to systematic errors.  Finally, the possibility remains that 
our estimates of the primordial abundances are correct and the D -- $^4$He 
tension is a hint of ``new physics".  For example, if the tau neutrino 
were massive ($\sim 5 - 20$ MeV) and unstable (lifetime $\sim 0.1 - 10$ 
sec.), the ``effective" number of equivalent light neutrinos would be 
less than the standard model case of N$_{\nu}$ = 3 (Kawasaki \etal 1994).  
For N$_{\nu} = 2.1 \pm 0.3$, consistency among the primordial abundances 
may be reestablished (Hata \etal 1995, Kawasaki, Kohri, \& Sato 1997).  
Other, non-standard, particle physics solutions are conceivable; degenerate 
neutrinos offer one such option (Kohri, Kawasaki, \& Sato 1997).

\section{Three Possible Resolutions of the Challenge to SBBN}

The simplest resolution of the SBBN crisis would be that we have been 
overly optimistic in believing the accuracy of our estimates of the 
primordial abundances derived from the observational data.  Primordial 
abundances are derived, not ``observed".  Each step in the process of 
acquiring, reducing, and analyzing the data presents opportunities for 
systematic as well as statistical errors to rear their ugly heads.  We 
are fortunate in that the various light element abundances are derived 
from very different observations (deuterium via absorption in the UV 
and optical parts of the spectrum from neutral interstellar gas and 
from counting particles in the solar wind and in lunar and meteoritic 
material; helium-4 from optical emission of hot, ionized gas in extragalactic 
\hii regions which have been relatively little polluted with the debris 
of stellar evolution; lithium via absorption in the spectra of very old 
stars which, as with the extragalactic \hii regions, were formed of very 
nearly primordial material).  As a result, it is very unlikely that errors 
in deriving the primordial abundance of one nuclide have any connection 
with those associated with the inferred abundances of the others.  For this 
reason I will review the current status of the abundance determinations, 
along with their associated uncertainties, separately for deuterium and 
for helium-4.

\subsection{Deuterium: Current Status}
In the best of all worlds we would prefer observations of high-redshift
regions (\ie, early in the evolution of the Universe) which offer direct
evidence of very little pollution by stellar evolution (low abundances 
of the ``metals", the heavy elements cooked in stars).  In the last few 
years just such data have become available for deuterium as a result of 
the Keck telescope observations of high-redshift (hi-z), low-metallicity 
(lo-Z) absorbers lying between us and very distant QSOs.  Unfortunately, 
the early data from these observations appear inconsistent and contradictory
(Songaila \etal 1994; Carswell \etal 1994; Rugers \& Hogan 1996; Tytler, 
Fan, \& Burles 1996; Burles \& Tytler 1997a).  On the one hand, the earliest 
data argue for a very high D abundance which, while consistent with the 
SBBN prediction if the inferred helium abundance were correct, seems 
inconsistent with the ISM and solar system data (Steigman 1994; Steigman 
\& Tosi 1995; Dearborn, Steigman, \& Tosi 1996; Tosi \etal 1998).  The 
impact of these data is shown on Figure 3 (labelled ``High D$_{\rm QSO}$"), 
the analog of Figure 2 with the hi-z, lo-Z QSO absorption data supplementing 
the solar system and ISM estimates of primordial deuterium.  As is clear 
from Figure 3, such a high primordial deuterium abundance relieves the 
tension between the predictions of SBBN and the observational data.  
However, if such a high primordial D abundance is correct a new challenge 
is posed, not to SBBN but to our understanding of the evolution of the 
Galaxy.  How is it that the Galaxy has managed to destroy deuterium by 
an order of magnitude while keeping the gas fraction relatively large 
and the heavy element abundance relatively small?  After all, if deuterium 
in the ISM has been destroyed by an order of magnitude then 90\% of the 
ISM has been cycled through stars, so why is it that more mass is not 
tied up in long-lived, low-mass stars and/or stellar remanants, and 
where are the metals produced by these stars?
\begin{figure}
\centerline{\psfig{file=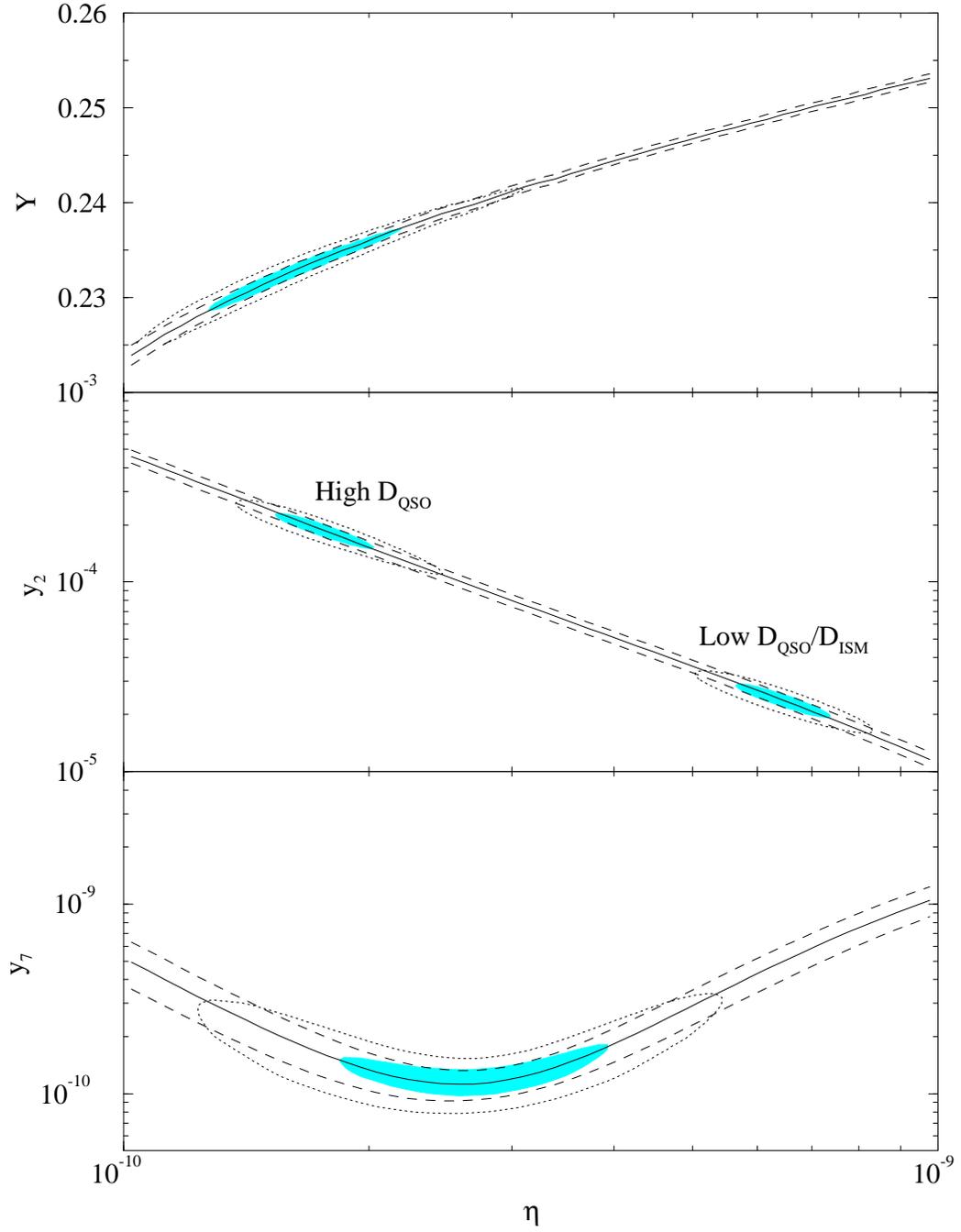,width=0.9\textwidth}}
\caption{As in Figure 2 with the high deuterium-abundance QSO measurements 
from Rugers \& Hogan (1996) and the low deuterium-abundance QSO data from 
Tytler, Fan, \& Burles (1996).  This figure is from Hata \etal (1997).}
\end{figure}

Hints that the early claims of high deuterium abundances may have been
contaminated by interloping hydrogen (Steigman 1994) receive support 
from subsequent Keck observations (Tytler, Fan, \& Burles 1996; Burles 
\& Tytler 1997a,b,c,d).  The implications for SBBN of these data are also 
shown in Figure 3 (labelled ``Low D$_{\rm QSO}$ /D$_{\rm ISM}$").  Although 
entirely consistent with ISM and solar system data and the evolution of 
the Galaxy (\eg, Tosi \etal 1998), these low D abundances exacerbate the 
crisis with SBBN.  At present, the observational situation is unclear.  
While some of the early claims of high-D have been withdrawn or modified 
(Hogan 1997), others persist and have been supplemented with new data 
from the Hubble Space Telescope (Webb \etal 1997).  At the same time, 
there has been some rachetting up of the low D abundances reported earlier; 
the current best estimate from Burles \& Tytler (1997d) is (D/H)$_{\rm P} 
= 3.40 \pm 0.25 \times 10^{-5}$ which, in SBBN, corresponds to $\eta_{10} 
= 5.1 \pm 0.3$.  This range of $\eta$ is just barely consistent with the 
lithium abundance (see, \eg, Pinsonneault \etal 1998) but is inconsistent 
with a low helium abundance (see Fig. 3).  It is to be hoped that more 
hi-z/lo-Z data will help resolve the hi-D/lo-D paradox.  Until then, it 
is important to consider the uncertainties in the derived primordial 
abundance of $^4$He.

\subsection{Helium-4: Current Status}

Helium-4 is a key player in testing the consistency of SBBN.  As the 
second most abundance element (after hydrogen) in the Universe, it may 
be observed throughout the Universe and its abundance determined to a 
much higher statistical accuracy than that of any other element.  As 
the most tightly bound of the light nuclides, $^4$He captures most of 
the neutrons which were present when primordial alchemy began so that 
its predicted primordial abundance is relatively insensitive to $\eta$.  
But since the neutron abundance at BBN depends on the competition between 
the weak interaction rate (interconverting neutrons and protons) and 
the universal expansion rate (driven by the total energy density which, 
at that time, is dominated by relativistic particles), the helium abundance 
provides a probe of the early expansion rate and, indirectly, of the 
particle content of the early Universe (Steigman, Schramm, \& Gunn 1977).  
It is the relative insensitivity of Y$_{\rm P}$ to $\eta$ which elevates 
helium to such a key role in testing SBBN.  Either the derived primordial 
abundance of helium does, or does not, agree with the SBBN prediction 
based on consistency with the abundances of the other light nuclides.  
For example, a ``high" deuterium abundance (D/H)$_{\rm P} = 2.0\times
10^{-4}$ corresponds to $\eta_{10} = 1.7$ and a predicted primordial 
helium mass fraction Y$_{\rm P} = 0.234$, while a ``low" deuterium 
abundance (D/H)$_{\rm P} = 3.4\times 10^{-5}$ favors $\eta_{10} = 5.1$ 
and Y$_{\rm P} = 0.247$.  If Y$_{\rm P}$ can be determined to better 
than $\pm 0.01$, the SBBN crisis can be resolved or confirmed.

With data assembled from the literature, Olive \& Steigman (1995) 
utilized the helium abundances derived from observations of low-metallicity,
extragalactic \hii regions in an attempt to pin down the primordial
helium abundance.  The restriction to low-metallicity regions is to
minimize the contamination from stellar-produced helium.  Nonetheless,
Olive \& Steigman (1995) find that even these low-metallicity data are 
correlated in the sense that, on average, the helium abundance increases 
with metallicity.  From a linear fit of helium (mass fraction) to oxygen 
abundance, Olive \& Steigman (1995) derived Y$_{\rm P} = 0.232 \pm 0.003$, 
leading to a $2\sigma$ upper bound Y$_{\rm P} \leq 0.238$.  With allowance 
for unknown systematic uncertainties of order 0.005, this suggests a 
robust upper bound to primordial helium of Y$_{\rm P}^{\rm MAX} \leq 
0.243$.  This estimate is consistent with ``high" deuterium but not with 
``low" deuterium (see Fig. 3).  However, new \hii region data (Izotov, 
Thuan \& Lipovetsky 1994) suggests a higher primordial abundance of 
helium, more consistent with the ``low" deuterium value.  Using all the 
helium data which had become available, Olive, Skillman \& Steigman (1997) 
found that the newer data were entirely consistent with previous observations 
and they derived for the combined data set (62 \hii regions) Y$_{\rm P} 
= 0.234 \pm 0.002$.  Thus, unless there are presently unidentified, 
large systematic offsets, consistency between D and $^4$He requires 
high primordial deuterium (low $\eta$), in conflict with some hi-z, lo-Z 
determinations (Burles \& Tytler 1997a,b,c,d) and with solar system and 
local ISM data (Steigman \& Tosi 1992, 1995; Dearborn, Steigman \& Tosi 
1996; Tosi \etal 1998).  

More recently, Izotov \& Thuan (1997) have added substantially to their 
independent data set of helium abundances in low metallicity \hii regions.  
As they have accumulated more data, the gap between their abundances and 
those from earlier work has grown.  The Izotov \& Thuan (1997) estimate 
of Y$_{\rm P} = 0.244 \pm 0.002$ differs by some 4-5$\sigma$ from the 
Olive \& Steigman (1995) value (note that the somewhat higher estimate 
of Olive, Skillman \& Steigman (1997) already included many of the Izotov 
\& Thuan (1997) regions).  There is apparently a systematic offset which 
is much larger than the identified statistical errors.  Although some of 
the difference is traceable to differing corrections for unseen neutral 
helium ($\lsim 0.002$) and some to a different approach to correcting 
for collisional excitation ($\lsim 0.002$), nonetheless an unexplained 
offset ($\gsim 0.008$) remains.  As is clear from Figure 3, the Izotov 
\& Thuan (1997) results are fully consistent with ``low" deuterium inferred 
from solar system and local ISM data as well as from some QSO absorption 
line systems.  However, until the offset in helium abundance determinations 
is understood it may be premature to heave a sigh of relief and declare 
the ``crisis" resolved.  Therefore, it remains of interest to investigate 
the consequences of the possibility that {\bf both} the deuterium and 
helium abundances may be ``low", and that the SBBN crisis is due to a 
small perturbation in the early expansion rate of the Universe traceable, 
perhaps, to non-standard particle physics (``beyond the standard model").

\subsection{Neutrino Counting}
The expansion of the Universe is driven by the energy density of its
constituents.  In the early Universe, at the time of primordial alchemy,
the energy density is dominated by ``radiation", the relativistic particles
(and antiparticles) present in the thermal bath.  In SBBN these are photons,
electron-positron pairs, and three families of light (\ie, relativistic)
neutrinos. If $\rho_{\gamma}$, $\rho_{e^{\pm}}$, and $\rho_{\nu}$ are the
energy densities for photons, electron-positron pairs, and one species of
light (massless) neutrinos, the total energy density may be written as,

\begin{equation}
\rho_{\rm TOT}^{\rm BBN} = \rho_{\gamma} + \rho_{e^{\pm}} + N_{\nu}\rho_{\nu}
\end{equation}
For SBBN, the ``effective" number of (equivalent) light neutrinos is the
standard model value $N_{\nu} = 3$.  If, however, the energy density of 
the early Universe (at a fixed temperature) differs from its standard 
value, the Universe will expand either faster or slower, leaving less or 
more time for neutrons to decay to protons and for nuclear reactions to 
occur.  This effect may be parameterized by $N_{\nu}$; if $N_{\nu} > 3$, 
the Universe expands faster, leaving more neutrons which results in more 
helium being synthesized.  As already noted, the D -- $^4$He data suggest
the oppposite, $N_{\nu} = 2.1 \pm 0.3$ (Hata \etal 1995).

Neutrino counting in the Big Bang differs in some crucial ways from 
neutrino counting at LEP.  For BBN, $N_{\nu}$ measures the contribution 
to the total energy density in the early Universe of any new particles 
(\eg, light scalars, sterile neutrinos, etc.) or of new physics (\eg,
degenerate neutrinos).  In contrast, experiments at particle accelerators 
such as LEP are sensitive to particles which possess full (or nearly full) 
strength weak interactions, even if they are very massive 
($\lsim M_{\rm Z}/2$).  For example, as Kawasaki \etal (1994) have shown, 
if the tau-neutrino were very massive but unstable, decaying into a 
mu-neutrino and a light, weakly-interacting scalar, $N_{\nu} < 3$ during 
BBN is possible.  Such a massive tau-neutrino would be ``light" at LEP 
and would be counted there as one massless neutrino species.  As Kawasaki, 
Kohri \& Sato (1997) have shown (see also Kawasaki \etal 1994), the D -- 
$^4$He tension would be relieved if the tau-neutrino had a mass in the 
range 10 -- 20 MeV and a lifetime within an order of magnitude of 0.1 
second.  Alternatively, Kohri, Kawasaki, \& Sato (1997) have shown that 
neutrino degeneracy (excess of neutrinos over antineutrinos, or vice-versa) 
may also resolve the crisis for SBBN.  Although it is difficult to conceive 
of testing the latter proposal, current accelerator experiments do have 
the potential to probe tau-neutrino masses in the ``interesting" range 
identified above.

\section{Summary}

The predictions of SBBN are observationally challenged.  The primordial
abundances of D and $^4$He inferred from observational data appear to 
be inconsistent with the predictions of SBBN.  Several options present 
themselves with the potential to resolve this crisis.  Perhaps the data
are at fault.  The conflicting deuterium abundances derived from observations
of high-redshift, low-metallicity QSO absorbers point an incriminating
finger.  If these data are supplemented with solar system and ISM deuterium
abundances, the lower D/H ratios are preferred.  But, is the extrapolation
from here and now (solar system, ISM) to there and then (primordial) under
control, or might there be unidentified systematic errors lurking?  The
two sets of apparently inconsistent helium abundances suggest systematic
errors at play in the extragalactic \hii region abundance determinations.
Although new data is always welcome, it is clear that a better understanding
of existing data may prove even more important.

In the absence of new data and/or a better understanding of the extant 
data it may be worthwhile to look elsewhere for clues.  My colleagues 
and I (Steigman, Hata, \& Felten 1997; SHF) have discarded the constraint 
on $\eta$ from SBBN and have utilized four other observational constraints
(Hubble parameter, age of the Universe, cluster gas (baryon) fraction, and 
effective ``shape" parameter $\Gamma$) to predict the three key cosmological 
parameters (Hubble parameter, total matter density, and the baryon density 
or $\eta$).  Considering both open and flat CDM models and flat $\Lambda$CDM 
models, SHF tested goodness of fit and drew confidence regions by the 
$\Delta\chi^2$ method.  In all of these models SHF find that large 
$\eta_{10}$ ($\gsim~6$) is favored strongly over small $\eta_{10}$ ($\lsim~2$), 
supporting reports of low deuterium abundances on some QSO lines of sight, 
and suggesting that observational determinations of primordial $^4$He may be 
contaminated by systematic errors.  

\section*{Acknowledgments}
I am indebted to my many collaborators whose joint work I have presented 
here.  Special thanks for their advice and assistance are due J. E. Felten, 
N. Hata, K. A. Olive, M. Pinsonneault, D. Thomas, M. Tosi, and T. P. Walker.  
This research is supported at Ohio State by DOE grant DE-FG02-91ER-40690.
  
\newcommand{\apj}{{\em Ap. J.}}
\newcommand{\apjs}{{\em Ap. J. Suppl.}}
\newcommand{\aj}{{\em A. J.}}
\newcommand{\mnras}{{\em MNRAS}}
\newcommand{\apjl}{{\em Ap. J. Lett.}}
\newcommand{\nat}{{\em Nature}}
\newcommand{\pl}{{\em Phys. Lett.}}

\end{document}